\documentclass[fleqn,11pt]{article}
\usepackage{hyperref}
\usepackage{amsmath, amssymb, multirow, graphicx,footnote,caption}
\usepackage{color}
\newtheorem{algorithm}{Algorithm}[section]

\begin{document}

\title{\bf{A New Estimation Algorithm for Box-Cox Transformation Cure Rate Model and Comparison With EM Algorithm}}
\author{\bf{Suvra Pal${}^{1}$\footnote{Corresponding author. E-mail address: suvra.pal@uta.edu Tel.: 817-272-7163.\newline \indent . } and Souvik Roy${}^{1}$}\\\\
${}^{1}$Department of Mathematics, University of Texas at Arlington, TX, 76019, USA.}
\date{}

\maketitle

\begin{abstract}

\noindent In this paper, we develop a new estimation procedure based on the non-linear conjugate gradient (NCG) algorithm for the Box-Cox transformation cure rate model. We compare the performance of the NCG algorithm with the well-known expectation maximization (EM) algorithm through a simulation study and show the advantages of the NCG algorithm over the EM algorithm. In particular, we show that the NCG algorithm allows simultaneous maximization of all model parameters when the likelihood surface is flat with respect to a Box-Cox model parameter. This is a big advantage over the EM algorithm, where a profile likelihood approach has been proposed in the literature that may not provide satisfactory results. We finally use the NCG algorithm to analyze a well-known melanoma data and show that it results in a better fit. 


\end{abstract}


\section{Introduction}

Advancements in the treatment of certain types of diseases (e.g., cancer, heart disease, etc.) has led to a noteworthy number of patients to respond favorably to the treatment, thereby not showing recurrence until the end of a long follow-up time. These patients are termed as recurrence-free survivors. It is quite possible that some of these patients will not show recurrence for a decently long period after the follow-up time since they may have reached a stage where the disease is undetectable and harmless. These patients, among the recurrence-free survivors, are termed as long-term survivors or ``cured''. The estimation of the proportion of cured patients cannot be readily obtained from a given lifetime data since it is not possible to identify which of the recurrence-free survivors can be considered as long-term survivors. This is because a patient who is susceptible to disease recurrence soon after the follow-up time may also show no recurrence and survive until the end of the follow-up time. The estimation of a treatment-specific cured proportion (or cure rate), however, is important to see the trend in the survival of patients suffering from a particular disease. It is also an important measure to judge the treatment's efficacy and its adoption in practice, as opposed to the standard treatment.

The first studied cure rate model dates back to the works of Boag (1949) and Berkson and Gage (1952), which is known as the mixture cure rate model. According to the mixture cure rate model, the population survival function of a time-to-event variable $Y$ is given by 
\begin{equation}
S_p(y) = p_0 + (1-p_0)S_s(y),
\label{mix}
\end{equation}
where $p_0$ is the cure rate and $S_s(\cdot)$ is the survival function of the susceptible patients only (a proper survival function). To incorporate a competing risks scenario, where several risk factors compete to produce the event of interest, Chen et al. (1999) proposed the promotion time cure rate model. In this case, the population survival function is given by 
\begin{equation}
S_p(y) = e^{-\eta(1-S(y))},
\label{prom}
\end{equation}
where $\eta$ represents the mean number of competing risks and $S(\cdot)$ represents the common survival function corresponding to the time taken by each risk factor to produce the event. Note that the cure rate is given by $e^{-\eta}$. To bring in the effect of prognostic factors on the cure rate, we may relate $p_0$ in \eqref{mix} to the prognostic factors using a logistic link function for the mixture cure rate model; see Kannan et al. (2010). Similarly, for the promotion time cure rate model, we may relate $\eta$ to the prognostic factors using a log-linear link function; see Rodrigues et al. (2009). Several approaches have been proposed in the literature to develop the associated inference for some flexible cure rate models that include the aforementioned mixture and promotion time cure rate models. In this regard, interested readers may refer to parametric approaches (Balakrishnan and Pal, 2013; Farewell, 1986; deFreitas and Rodrigues, 2013; Balakrishnan and Pal, 2016); semi-parametric approaches (Kuk and Chen, 1992; Li and Taylor, 2002; Balakrishnan et al., 2017); and non-parametric approaches (Balakrishnan et al., 2016; Maller and Zhou, 1996).

In this paper, we consider a unification of the mixture and promotion time cure rate models through a novel class of cure rate model based on the Box-Cox transformation (BCT) (Box and Cox, 1964) on the population survival function, which was originally proposed by Yin and Ibrahim (2005). The BCT cure rate model was very recently studied in Pal and Balakrishnan (2018), where the authors developed the expectation maximization (EM) algorithm for the maximum likelihood estimation (MLE) of the model parameters. It was noted by the authors that the likelihood surface was flat with respect to the Box-Cox index parameter $\alpha$ (using the same notation as in Pal and Balakrishnan, 2018). As such, simultaneous maximization of all model parameters was not possible. To circumvent this problem, the authors proposed a profile likelihood approach within the EM algorithm with respect to the index parameter $\alpha$. Although the profile likelihood approach worked well in retrieving the true value of $\alpha$, few drawbacks were noted. For instance, the mean square error of $\alpha$ as well as the regression parameters were pretty high compared to the other model parameters; see Table III in Pal and Balakrishnan (2018). Furthermore, using the profile likelihood approach, the EM algorithm had to be run for each chosen fixed value of $\alpha$. To evade these issues with the EM algorithm, we propose a new estimation procedure based on the non-linear conjugate gradient (NCG) algorithm that allows simultaneous maximization of all model parameters and, at the same time, provides satisfactory estimation results. We show the advantages of the NCG algorithm over the well-known EM algorithm through a detailed simulation study.

The rest of the paper is organized as follows. In Section II, we introduce the BCT cure rate model. In Section III, we discuss the form of data and likelihood function. In Section IV, we develop the NCG algorithm for the BCT model. In Section V, we discuss a simulation study and compare the model fitting results of the NCG algorithm with the EM algorithm. In Section VI, we apply the NCG algorithm to analyze a well-known melanoma data. Finally, in Section VII, we make some concluding remarks and present some future research in this direction.

\section{Box-Cox transformation cure rate model}

The BCT on a variable $z$ with index parameter $\alpha$ is given by; see Box and Cox (1964)
\begin{equation}
G(z,\alpha)=\begin{cases}
\frac{z^\alpha -1}{\alpha}, & 0<\alpha\leq 1\\
\log(z), & \alpha = 0.
\end{cases}
\label{BC} 
\end{equation}  
According to Yin and Ibrahim (2005), using the BCT on the population survival function, the BCT cure rate model can be defined as
\begin{equation}
G(S_p(y|\boldsymbol x),\alpha) = -\phi(\alpha,\boldsymbol x)F(y), \ \ 0\leq\alpha \leq 1.
\label{G}
\end{equation}
In \eqref{G}, $F(\cdot)$ is a proper distribution function and $\boldsymbol x$ is a set of prognostic factors introduced in the model through a general covariate structure $\phi(\alpha,\boldsymbol x)$, defined as
\begin{equation}
\phi(\alpha,\boldsymbol x) = \begin{cases}
\frac{\exp(\boldsymbol x^\prime\boldsymbol\beta)}{1+\alpha\exp(\boldsymbol x^\prime\boldsymbol\beta)}, & \ \ \ \ 0<\alpha\leq 1 \\ 
\exp(\boldsymbol x^\prime\boldsymbol\beta), & \ \ \ \ \alpha = 0
\end{cases}
\label{phi}
\end{equation}
with $\boldsymbol\beta$ denoting the set of regression coefficients. Using \eqref{BC} in \eqref{G}, we can easily obtain an expression for the population survival function as
\begin{equation}
S_p(y|\boldsymbol x) = \begin{cases}
\{1-\alpha\phi(\alpha,\boldsymbol x)F(y)\}^{\frac{1}{\alpha}}, & 0 < \alpha \leq 1\\
\exp\{-\phi(0,\boldsymbol x)F(y)\}, & \alpha = 0.
\end{cases}
\label{Sp}
\end{equation}
Note that the cure rate being the long-term survival probability can be obtained as $p_0(\boldsymbol x)=\lim_{y\rightarrow\infty}S_p(y|\boldsymbol x)$. Its explicit expression is given by
\begin{eqnarray}
p_0(\boldsymbol x)&=&\begin{cases}
[1-\alpha\phi(\alpha,\boldsymbol x)]^{\frac{1}{\alpha}}, & 0 < \alpha \leq 1\\
\exp\{-\phi(0,\boldsymbol x)\}, & \alpha = 0.
\end{cases}
\label{p0}
\end{eqnarray}
From \eqref{Sp}, we can derive the expression of the population density function as $f_p(y|\boldsymbol x) = S_p^{\prime}(y|\boldsymbol x)$. Its explicit expression is given by; see Pal and Balakrishnan (2018)
\begin{eqnarray}
f_p(y|\boldsymbol x) = \begin{cases}
S_p(y|\boldsymbol x)\phi(\alpha,\boldsymbol x)f(y)\\
\times \{1-\alpha\phi(\alpha,\boldsymbol x)F(y)\}^{-1}, & 0 < \alpha \leq 1\\ 
S_p(y|\boldsymbol x)\phi(0,\boldsymbol x)f(y), & \alpha = 0,
\end{cases}
\label{fp}
\end{eqnarray} 
where $f(\cdot)$ is the density function corresponding to $F(\cdot)$.

It is interesting to note that the BCT cure rate model in \eqref{Sp} reduces to the mixture cure rate model in \eqref{mix} with $p_0=p_0(\boldsymbol x)=\frac{1}{1+\exp(\boldsymbol x^\prime\boldsymbol\beta)}$ and $S_s(y) = 1-F(y)$. Furthermore, the BCT cure rate model reduces to the promotion time cure rate model in \eqref{prom} with $S(y)=1-F(y)$ and $\eta=\exp(\boldsymbol x^\prime\boldsymbol\beta)$. Thus, the BCT cure rate model unifies the two most commonly used cure rate models in the literature and provides a flexible class of transformation cure rate model. Although the BCT cure rate model has not been extensively studied in the literature, few authors have considered it. For instance, a novel biological interpretation of the BCT model was addressed in Peng and Xu (2012). In the context of frailty model with multivariate survival data, Diao and Yin (2012) considered the BCT model and developed non-parametric estimation procedure. Interested readers may also refer to Zeng et al. (2006), Sun et al. (2011), Castro et al. (2010), and Koutras and Milienos (2017).

\section{Form of data and likelihood function}

We work with a scenario where the time-to-event data may not be completely observed and is subject to right censoring. Let $T_i$ and $C_i$ denote the actual failure time and censoring time, respectively, for $i=1,2,\cdots,n$ with $n$ denoting the sample size. The observed lifetime is then given by $Y_i=\min\{T_i,C_i\}$. Let $\delta_i$ denote the right censoring indicator which takes the value 1 if the time-to-event is observed and 0 if it is right censored. The set of observed data can thus be given by $\boldsymbol O = \{(y_i,\delta_i,\boldsymbol x_i), i=1,2,\cdots,n\}$, where $\boldsymbol x_i$ is the vector of prognostic factors corresponding to the $i$-th subject. Assuming non-informative censoring, we can define the observed data likelihood function as 
\begin{eqnarray}
L(\boldsymbol\theta)= \prod_{i=1}^n \{f_p(y_i|\boldsymbol {x_i})\}^{\delta_i}\{S_p(y_i|\boldsymbol{x_i})\}^{1-\delta_i}, \label{eq:likelihood}
\end{eqnarray}
where $\boldsymbol{\theta}=(\alpha,\boldsymbol\beta^\prime,\boldsymbol\gamma^\prime)^\prime$ is the vector of unknown model parameter with $\boldsymbol\gamma$ denoting the parameter vector associated with $F(\cdot)$ in \eqref{Sp}.

As done in Pal and Balakrishnan (2018), we shall now assume $F(\cdot)$ in \eqref{Sp} and $f(\cdot)$ in \eqref{fp} to be the distribution function and density function, respectively, of a two-parameter Weibull distribution. Thus, we have 
\begin{eqnarray}
F(y)&=&1-\exp\{-(\gamma_2 y)^{\frac{1}{\gamma_1}}\} \ \ \text{and} \nonumber \\ 
f(y) &=& \frac{1}{\gamma_1 y}(\gamma_2 y)^{\frac{1}{\gamma_1}}\{1-F(y)\} 
\label{wei}
\end{eqnarray}
for $y>0$, $\gamma_1>0,$ and $\gamma_2>0.$ Also, we now have $\boldsymbol\gamma = (\gamma_1,\gamma_2)^\prime$.

\section{Estimation method: Non-linear Conjugate gradient algorithm}
In order to estimate the optimal parameter set $\boldsymbol\theta$ that maximizes the likelihood function given in (\ref{eq:likelihood}), we first apply the natural logarithm on both sides of (\ref{eq:likelihood}) to obtain the log-likelihood function as follows
\begin{equation}\label{eq:loglikelihood}
l(\boldsymbol\theta)= \sum_{i=1}^n[ \delta_i \log\{f_p(y_i|\boldsymbol {x_i})\}+{(1-\delta_i)}\log\{S_p(y_i|\boldsymbol{x_i})\}].
\end{equation}
Our maximization problem can now be stated as follows
\begin{equation}\label{eq:maxproblem}
\max_{\boldsymbol\theta\in U}~ l(\boldsymbol\theta),\\
\end{equation}
where the feasible set of constraints $U$ is defined as
\[
U = \lbrace \boldsymbol\theta : \gamma_1 > 0,\gamma_2 >0, ~ 0< \alpha \leq 1 \rbrace.
\]
The function $l$ depends non-linearly on $\boldsymbol\theta$, thus, rendering the above maximization problem non-linear. Thus, to solve the maximization problem (\ref{eq:maxproblem}), we use the non-linear conjugate gradient method (NCG), which is an extension of the traditional conjugate gradient method for a linear optimization problem; see Neittaanmaki and Tiba (1994). Such a scheme has been used to solve partial differential equation (PDE)-constrained optimal control problems arising in mathematical models of crowd motion, game theory and medical imaging (Roy et al., 2016; Roy et al., 2017; Roy et al., 2018; Adesokan et al., 2018). A major advantage of using the NCG scheme over traditional Newton-based schemes (for instance, the EM algorithm in Pal and Balakrishnan, 2018 where the maximization step is carried out using a one-step Newton Raphson method) is that even though Newton's method potentially converges faster, the computation of the Hessian in a Newton-based method is highly expensive and time consuming, specially for optimization problems with large number of parameters and bigger sample size $n$. On contrary, one requires to compute only the gradient in the NCG method and, thus, for such problems, the NCG scheme results in a faster convergence than the Newton-based schemes. This is also observed in the simulation studies presented in Section \ref{sec:simulation}.

In order to describe this method, we start off with an initial guess $\boldsymbol\theta_0$ for the parameter set. We now update this initial guess by moving in the search direction given by the gradient $\boldsymbol{d}_0=\boldsymbol{g}_0=\nabla_{\boldsymbol\theta} l(\boldsymbol\theta_0)$ of the function $l$. The rationale behind this kind of an update arises from the fact that the maximum rate of increase of a function is along the positive gradient direction. In subsequent iterations, the search directions are recursively given by the formula 
\[
\boldsymbol{d}_{k+1} = \boldsymbol{g}_{k+1}+\xi_k \boldsymbol{d}_k,~ k=0,1,2,\cdots,
\]
where $\boldsymbol{g}_k = \nabla_{\boldsymbol\theta} l(\boldsymbol\theta_k)$ and $\xi_k$ is given by the formula of Hager-Zhang (Hager and Zhang, 2005) as follows
\begin{equation}\label{eq:step_HG}
\xi_k = \dfrac{1}{\boldsymbol{d}_k' \boldsymbol{w}_k}\left(\boldsymbol{w}_k-2\boldsymbol{d}_k\dfrac{\boldsymbol{w}_k' \boldsymbol{w}_k}{\boldsymbol{d}_k' \boldsymbol{w}_k}\right)' \boldsymbol{g}_{k+1}
\end{equation}
with $\boldsymbol{w}_k = \boldsymbol{g}_{k+1}-\boldsymbol{g}_k$. We update our parameter set $\boldsymbol\theta$ using a steepest ascent scheme given below
\[
\boldsymbol\theta_{k+1} = \boldsymbol\theta_k+s_k \boldsymbol{d}_k,
\]
where $s_k>0$ is a steplength obtained through a line search algorithm that uses the following Armijo condition (Annunziato and Borzi, 2013) of sufficient increase of $l$
\[
l(\boldsymbol\theta_k + s_k \boldsymbol{d}_k) \geq l(\boldsymbol\theta_k) + \lambda ~\boldsymbol{d}_k' \boldsymbol{g}_k
\]
with $0 < \lambda < 1/2$. Further, a projection step onto the constraint set $U$ is applied to this parameter update step through the following way
\[
\boldsymbol\theta_{k+1} = \mathbb{P}[\boldsymbol\theta_k+s_k \boldsymbol{d}_k],
\]
where 
\[
\begin{aligned}
\mathbb{P}[\boldsymbol\theta] = &\lbrace \max(10^{-10}, \min(1, \alpha)),\boldsymbol\beta, \\
&\max(10^{-10},\gamma_1),\max(10^{-10},\gamma_2)\rbrace.
\end{aligned}
\]
The projection step ensures that the parameter values obtained in each iteration lies inside the constraint set $U$. The projected NCG algorithm is terminated once the relative difference between two successive iterates is less than a specified tolerance level or the number of iterations exceed the maximum number of iterations. The algorithm is summarized below.
\begin{algorithm}[Projected NCG Scheme]\label{alg:NCG}
\begin{enumerate}
\item Input: initial guess $\boldsymbol\theta_0$. Evaluate $\boldsymbol{d}_0 = \nabla_{\boldsymbol\theta}l(\boldsymbol\theta_0)$, index $k=0$, maximum $k=k_{max}$, tolerance = $tol$
\item While $(k<k_{max}),$ do
\item Set $\boldsymbol\theta_{k+1} = \mathbb{P}[\boldsymbol\theta_k + s_k\boldsymbol{d}_k]$, where $s_k$ is obtained using the line-search algorithm
\item Compute $\boldsymbol{g}_{k+1} = \nabla_{\boldsymbol\theta}l(\boldsymbol\theta_{k+1})$
\item Compute $\xi_k$ using (\ref{eq:step_HG})
\item Set $\boldsymbol{d}_{k+1}=\boldsymbol{g}_{k+1}+\xi_k \boldsymbol{d}_k$
\item If $\|\frac{\boldsymbol\theta_{k+1}-\boldsymbol\theta_k}{\boldsymbol\theta_k}\| < tol$, terminate
\item Set $k=k+1$
\item End while.
\end{enumerate}
\end{algorithm}
The convergence of the projected NCG scheme, as described in Algorithm \ref{alg:NCG}, follows from Neittaanmaki and Tiba (1994, Lemma 1.6, p. 235).

\section{Simulation study} \label{sec:simulation}

To demonstrate the performance of our proposed NCG algorithm, we carried out a Monte Carlo simulation study. For this purpose, we chose $k_{max} = 100,$ $\lambda = 0.1$ (for other values of $\lambda$ within the range, our findings were similar),  and $tol = 0.001.$ To compare the performance of the NCG algorithm with the already developed EM algorithm (where the maximization step is done using a one-step Newton Raphson method) in Pal and Balakrishnan (2018), we stick to the same simulation setup as in Pal and Balakrishnan (2018). Below, we briefly describe the simulation setup considered by the authors in Pal and Balakrishnan (2018). All computational codes were developed using R statistical software.

\subsection{Using binary covariate} \label{sec:SA}

Here, we consider one binary covariate $x$, for instance, gender, ulceration status, etc. This is equivalent to assuming a scenario where patients are divided into two groups with $x=1$ denoting patients that belong to group 1 and $x=0$ denoting patients that belong to group 2. We denote the cured proportions corresponding to groups 1 and 2 by $p_{01}$ and $p_{00},$ respectively. On assuming true values for the cured proportions $p_{01}$ and $p_{00}$ and a chosen value for $\alpha$, we can calculate the true values of the regression parameters as; see Pal and Balakrishnan (2018)
 \begin{eqnarray*}
\beta_0& =& \begin{cases}
\log\bigg[\frac{1}{\alpha}\bigg\{\frac{1}{p_{00}^\alpha}-1\bigg\}\bigg], & 0 < \alpha \leq 1\\ 
\log\{-\log(p_{00})\}, & \alpha = 0 
\end{cases}
\end{eqnarray*} 
and
\begin{eqnarray*}
\beta_1&=&\begin{cases}
\log\bigg[\frac{1}{\alpha}\bigg\{\frac{1}{p^\alpha_{01}}-1\bigg\}\bigg]-\beta_0, & 0 < \alpha \leq 1\\ 
\log\{-\log(p_{01})\}-\beta_0, & \alpha = 0.
\end{cases}
\end{eqnarray*}

For a given group, if we denote the cured proportion by $p_0,$ we can use the following steps to generate data from the BCT model in~\eqref{Sp}
\begin{itemize}
\item[(i)] We first generate a Uniform(0,1) random variable ($U$) and a censoring time ($C$) from an exponential distribution with rate $c$. Note that the censoring rate can be different for different groups;
\item[(ii)] If $U\leq p_0,$ the observed time $Y$ is taken as the censoring time, i.e., $Y=C$;
\item[(iii)] If $U>p_0,$ we equate $\frac{S_p(y|\boldsymbol x)-p_0}{1-p_0}$ (survival probability of the susceptible patients) to $U^*,$ where $U^*\sim$ Uniform(0,1). Thus, we have
\begin{eqnarray*}
&&
\begin{cases}
\frac{\{1-\alpha\phi(\alpha, x)F(t)\}^{\frac{1}{\alpha}}-p_0}{1-p_0}= U^*, & 0 < \alpha \leq 1  \\\\
\frac{\exp\{-\phi(0,x)F(t)\}-p_0}{1-p_0}= U^*, & \alpha = 0
\end{cases}\\
&\Rightarrow&
\begin{cases}
t=F^{-1}\bigg[\frac{1-\{p_0+(1-p_0)U^*\}^\alpha}{\alpha\phi(\alpha, x)}\bigg], & 0 < \alpha \leq 1  \\\\
t=F^{-1}\bigg[\frac{-\log\{p_0+(1-p_0)U^*\}}{\phi(0, x)}\bigg], & \alpha = 0,
\end{cases}
\end{eqnarray*}
where $F$ is the distribution function of the Weibull distribution as given in \eqref{wei}. The observed time can then be obtained as $Y=\min\{t,C\}$;
\item[(iv)] From (ii) and (iii), if $Y=C,$ we set $\delta=0;$ otherwise, we set $\delta=1$.
\end{itemize}

We consider the same parameter settings as in Pal and Balakrishnan (2018), i.e., three different choices of sample sizes as $n=150$ $(n_1=75,n_2=75)$, $n=200$ $(n_1=120,n_2=80)$ and $300$ $(n_1=180,n_2=120)$; two different choices of cured proportions $(p_{01},p_{00})$ as $(0.65,0.35)$ and $(0.40,0.20)$; a particular choice of $(\gamma_1,\gamma_2)$ as (0.316,0.179); and a particular choice of censoring rates for groups (1,2) as (0.15, 0.10). For the initial values of model parameters, we follow the same procedure as in Pal and Balakrishnan (2018).

\subsection{Using continuous covariate}

Here, we consider a continuous covariate $x$, for instance, the thickness of tumor (in mm), which can be generated from a Uniform(0.1,20) distribution. Since we expect the cured proportion to decrease with an increase in tumor thickness, we can fix a low value of cured proportion as $p_{\text{low}} = 5\%$ for a patient with tumor thickness $x_{\text{max}}=$ 20mm and a high value of cured proportion as $p_{\text{high}}=65\%$ for a patient with tumor thickness $x_{\text{min}}=$ 0.1mm. Thus, we can calculate the true values of the regression parameters, for a fixed choice of $\alpha$, as; see Pal and Balakrishnan (2018)
 \begin{eqnarray*}
\beta_1& =& \begin{cases}
\frac{\log\bigg[\frac{1}{\alpha}\bigg\{\frac{1}{p_{\text{low}}^\alpha}-1\bigg\}\bigg] - \log\bigg[\frac{1}{\alpha}\bigg\{\frac{1}{p_{\text{high}}^\alpha}-1\bigg\}\bigg]}{x_{max}-x_{min}}, & 0 < \alpha \leq 1\\\\
\frac{\log\{-\log(p_{\text{low}})\}-\log\{-\log(p_{\text{high}})\}}{x_{max}-x_{min}}, & \alpha = 0 
\end{cases}
\end{eqnarray*} 
and
\begin{eqnarray*}
\beta_0&=&\begin{cases}
\log\bigg[\frac{1}{\alpha}\bigg\{\frac{1}{p^\alpha_{\text{low}}}-1\bigg\}\bigg]-\beta_1x_{max}, & 0 < \alpha \leq 1\\ 
\log\{-\log(p_{\text{low}})\}-\beta_1x_{max}, & \alpha = 0.
\end{cases}
\end{eqnarray*}

For a patient with tumor thickness $x$ and for a chosen true value of $\alpha$, we can use the following steps to generate data from the BCT model in~\eqref{Sp}
\begin{itemize}
\item[(i)] We first calculate the cure rate as
\begin{eqnarray*}
p_0(x)&=&
\begin{cases}
\bigg[\frac{1}{1+\alpha\exp(\beta_0+\beta_1x)}\bigg]^{\frac{1}{\alpha}}, & 0 < \alpha \leq 1\\
\exp(-\exp(\beta_0+\beta_1x)), & \alpha = 0;
\end{cases}
\end{eqnarray*}
\item[(ii)] We then generate $U^*\sim$ Uniform(0,1) and censoring time ($C$) from an exponential distribution with rate $c$;
\item[(iii)] If $U^*\leq p_0(x)$, the observed time $Y$ is taken as the censoring time, i.e., $Y=C$;
\item[(iv)] If $U^* > p_0(x)$, we follow step (iii) of section \ref{sec:SA} with $p_0$ replaced by $p_0(x)$;
\item[(v)] From (iii) and (iv), if $Y=C,$ we set $\delta=0$; otherwise, we set $\delta=1.$
\end{itemize}
In this study, we use two different sample sizes as $n=150$ and $n=200$. We keep the true values of lifetime parameters and the technique of finding initial values the same as in the simulation setup with binary covariate.

\subsection{Discussion}


In Table \ref{table:NCG}, we present the simulation study results, in terms of bias and root mean square error (RMSE), for the general BCT model when the covariate under study is binary. When we employed the NCG algorithm, there was no issue with simultaneous maximization of all model parameter (including $\alpha$). This behavior of the NCG algorithm is unlike the EM algorithm where simultaneous maximization is not possible and one needs to keep the parameter $\alpha$ fixed for the iterative procedure to work. This is a huge advantage of the NCG algorithm over the EM algorithm. We note that the NCG algorithm converges to the true parameter values very accurately with small bias and RMSE. Furthermore, the large sample properties, i.e., the reduction of bias and RMSE with an increase in sample size, seems to be satisfied quite well. When compared with the EM algorithm, the NCG algorithm results in a significant reduction of bias and RMSE for the parameters $\beta_0,$ $\beta_1,$ and $\alpha$. This is another big advantage of NCG algorithm over the EM algorithm, noting that the cure rate is a pure function of these three parameters only. For the lifetime parameters $\gamma_1$ and $\gamma_2,$ the bias and RMSE of both algorithms are comparable. From these observations, it is clear that the overall performance of the NCG algorithm is significantly better than the EM algorithm and is thus preferred for the BCT model.  In Tables \ref{table:a1} and \ref{table:a0}, we present the results for mixture model ($\alpha=1$) and promotion time model ($\alpha=0$), respectively, where the estimation of $\alpha$ is not involved. The NCG algorithm once again results in significantly lower RMSE for the regression parameters when compared to the EM algorithm. In Table \ref{table:NCG2}, we present the results when the covariate under study is continuous and the observations are similar to that in Table \ref{table:NCG}. In Tables \ref{table:NCG}-\ref{table:NCG2}, the EM results are taken from Pal and Balakrishnan (2018).

We next turn our attention to the bias and RMSE of the cure rates and present these in Table \ref{table:cure} for the general BCT model with binary covariate only since for the continuous covariate the cure rate will be different for each subject. Although the cure rates are estimated unbiasedly for both NCG and EM algorithms, note once again the reduction in the RMSE when employing the NCG algorithm. The observations are similar for other parameter settings as well as for the mixture $(\alpha=1)$ and promotion time $(\alpha=0)$ models and are not presented here for the sake of brevity. Once again, the EM results in Table \ref{table:cure} are taken from Pal and Balakrishnan (2018).

It is also important to compare the CPU times between the NCG and EM algorithms. For this purpose, we considered the general BCT model and present the CPU times (in seconds) in Table \ref{table:CPU} for some parameter settings. Each reported time represents the total time taken by an algorithm to produce the estimation results (estimates, bias, and RMSE) over 500 Monte Carlo runs. We note that the NCG algorithm runs faster if the sample size is large or the cured proportions are low. The EM algorithm, on the other hand, does run faster for lower cured proportions, however, unlike the NCG algorithm, the EM algorithm takes more time to run for larger sample size. When comparing between the NCG and EM algorithms for a given parameter setting, the NCG algorithm runs faster if the sample size is large and this is true irrespective of the cured proportions. Having said this, we must mention that both algorithms produce estimation results in a reasonable amount of time.
\begin{table} [ht!]
\caption{Comparison of NCG algorithm with EM algorithm in terms of bias and RMSE for the BCT model with binary covariate}
\centering
\begin{tabular}{ l l l l l l l }\\ 
\hline                                 
$n$ & $(p_{01},p_{00})$ & Parameter &\multicolumn{2}{c}{Bias}  &\multicolumn{2}{c}{RMSE} \\  \cline{4-5}  \cline{6-7}
& & &NCG & EM & NCG & EM \\
\hline  

150 & (0.40,0.20)  & $\beta_0=0.905$                          & -0.014 & 0.113 & 0.160 & 0.547 \\
 & &  $\beta_{1}=-0.755$                                           & -0.011 & -0.076 & 0.214 & 0.518 \\
 & &  $\gamma_1=0.316$                                          & -0.007 & 0.000 & 0.032 & 0.032 \\ 
 & &  $\gamma_2=0.179$                                          & 0.001 & 0.003 & 0.013 & 0.014\\
 & &  $\alpha=0.500$                                                & 0.010  & 0.052 & 0.156 & 0.428\\[0.5ex]
 
200 & (0.40,0.20)  & $\beta_0=0.905$                         & -0.012 & 0.058 & 0.119 & 0.517 \\
 & &  $\beta_{1}=-0.755$                                           & -0.006 & -0.054 & 0.120 & 0.432 \\
 & &  $\gamma_1=0.316$                                          & -0.011 & 0.000 & 0.031 & 0.029 \\ 
 & &  $\gamma_2=0.179$                                          & 0.001 & 0.001 & 0.015 & 0.013\\
 & &  $\alpha=0.500$                                                & 0.006  & 0.007 & 0.102 & 0.418\\[0.5ex]
 
 300 & (0.40,0.20)  & $\beta_0=0.905$                         & 0.004 & 0.087 & 0.107 & 0.485 \\
 & &  $\beta_{1}=-0.755$                                           & 0.002 & -0.064 & 0.090 & 0.388 \\
 & &  $\gamma_1=0.316$                                          & -0.003 & 0.000 & 0.034 & 0.025 \\ 
 & &  $\gamma_2=0.179$                                          & 0.001 & 0.002 & 0.019 & 0.011\\
 & &  $\alpha=0.500$                                                & 0.004  & 0.058 & 0.064 & 0.405\\[0.5ex]
 
 150 & (0.65,0.35)  & $\beta_0=0.468$                        & -0.017 & -0.086 & 0.179 & 0.389 \\
 & &  $\beta_{1}=-1.144$                                           & 0.009 & 0.043 & 0.265 & 0.465 \\
 & &  $\gamma_1=0.316$                                          & -0.004 & -0.008 & 0.038 & 0.036 \\ 
 & &  $\gamma_2=0.179$                                          & 0.001 & -0.002 & 0.011 & 0.012\\
 & &  $\alpha=0.750$                                                & -0.002  & -0.193 & 0.207 & 0.487\\[0.5ex]
 
  200 & (0.65,0.35)  & $\beta_0=0.468$                       & -0.023 & -0.054 & 0.150 & 0.392 \\
 & &  $\beta_{1}=-1.144$                                           & -0.007 & 0.013 & 0.212 & 0.426 \\
 & &  $\gamma_1=0.316$                                          & -0.010 & -0.005 & 0.034 & 0.033 \\ 
 & &  $\gamma_2=0.179$                                          & 0.001 & 0.000 & 0.011 & 0.011\\
 & &  $\alpha=0.750$                                                & 0.002  & -0.136 & 0.164 & 0.446\\[0.5ex]
 
  300 & (0.65,0.35)  & $\beta_0=0.468$                        & -0.020 & -0.082 & 0.083 & 0.361 \\
 & &  $\beta_{1}=-1.144$                                           & -0.009 & 0.044 & 0.134 & 0.368 \\
 & &  $\gamma_1=0.316$                                          & -0.010 & -0.006 & 0.031 & 0.028 \\ 
 & &  $\gamma_2=0.179$                                          & 0.001 & -0.001 & 0.015 & 0.009\\
 & &  $\alpha=0.750$                                                & 0.006  & -0.167 & 0.096 & 0.461\\
\hline
\end{tabular}
\label{table:NCG}
\end{table} 
\begin{table} [ht!]
\caption{Comparison of NCG algorithm with EM algorithm in terms of bias and RMSE for the BCT$(\alpha=1)$ model with binary covariate}
\centering
\begin{tabular}{ l l l l l l l }\\ 
\hline                                 
$n$ & $(p_{01},p_{00})$ & Parameter &\multicolumn{2}{c}{Bias}  &\multicolumn{2}{c}{RMSE} \\  \cline{4-5}  \cline{6-7}
& & &NCG & EM & NCG & EM \\
\hline  

300 & (0.65,0.35)  & $\beta_0=0.619$                         & -0.025 & 0.017 & 0.105 & 0.261 \\
 & &  $\beta_{1}=-1.238$                                           & -0.014 & -0.036 & 0.155 & 0.345 \\
 & &  $\gamma_1=0.316$                                          & -0.012 & -0.002 & 0.032 & 0.027 \\ 
 & &  $\gamma_2=0.179$                                          & 0.000 & 0.001 & 0.015 & 0.007\\[0.5ex]
 
 300 & (0.40,0.20)  & $\beta_0=1.386$                         & -0.006 & 0.058 & 0.153 & 0.339 \\
 & &  $\beta_{1}=-0.981$                                           & -0.007 & -0.057 & 0.115 & 0.429 \\
 & &  $\gamma_1=0.316$                                          & -0.006 & -0.002 & 0.035 & 0.023 \\ 
 & &  $\gamma_2=0.179$                                          & 0.001 & 0.000 & 0.020 & 0.006\\[0.5ex]
 
 150 & (0.40,0.20)  & $\beta_0=1.386$                        & -0.031 & 0.078 & 0.212 & 0.460 \\
 & &  $\beta_{1}=-0.981$                                           & -0.022 & -0.039 & 0.251 & 0.601 \\
 & &  $\gamma_1=0.316$                                          & -0.008 & -0.004 & 0.032 & 0.030 \\ 
 & &  $\gamma_2=0.179$                                          & 0.001 & 0.001 & 0.012 & 0.008\\
 \hline
\end{tabular}
\label{table:a1}
\end{table}
\begin{table} [ht!]
\caption{Comparison of NCG algorithm with EM algorithm in terms of bias and RMSE for the BCT$(\alpha=0)$ model with binary covariate}
\centering
\begin{tabular}{ l l l l l l l }\\ 
\hline                                 
$n$ & $(p_{01},p_{00})$ & Parameter &\multicolumn{2}{c}{Bias}  &\multicolumn{2}{c}{RMSE} \\  \cline{4-5}  \cline{6-7}
& & &NCG & EM & NCG & EM \\
\hline  

300 & (0.65,0.35)  & $\beta_0=0.049$                         & -0.020 & -0.001 & 0.080 & 0.162 \\
 & &  $\beta_{1}=-0.891$                                           & 0.002 & 0.003 & 0.130 & 0.235 \\
 & &  $\gamma_1=0.316$                                          & -0.010 & -0.003 & 0.031 & 0.027 \\ 
 & &  $\gamma_2=0.179$                                          & 0.002 & 0.000 & 0.014 & 0.008\\[0.5ex]
 
 300 & (0.40,0.20)  & $\beta_0=0.476$                         & -0.004 & 0.000 & 0.062 & 0.137 \\
 & &  $\beta_{1}=-0.563$                                           & -0.003 & -0.003 & 0.065 & 0.182 \\
 & &  $\gamma_1=0.316$                                          & -0.006 & -0.002 & 0.033 & 0.022 \\ 
 & &  $\gamma_2=0.179$                                          & 0.002 & 0.000 & 0.019 & 0.008\\[0.5ex]
 
 150 & (0.40,0.20)  & $\beta_0=0.476$                        & -0.035 & 0.019 & 0.155 & 0.182 \\
 & &  $\beta_{1}=-0.563$                                           & 0.012 & -0.022 & 0.203 & 0.265 \\
 & &  $\gamma_1=0.316$                                          & -0.003 & -0.003 & 0.031 & 0.030 \\ 
 & &  $\gamma_2=0.179$                                          & 0.001 & 0.000 & 0.013 & 0.011\\
 \hline
\end{tabular}
\label{table:a0}
\end{table}  
\begin{table} [ht!]
\caption{Comparison of NCG algorithm with EM algorithm in terms of bias and RMSE for the BCT model with continuous covariate}
\centering
\setlength{\tabcolsep}{10pt}
\begin{tabular}{ l l l l l l }\\ 
\hline                                 
$n$ & Parameter &\multicolumn{2}{c}{Bias}  &\multicolumn{2}{c}{RMSE} \\  \cline{3-4}  \cline{5-6}
& &NCG & EM & NCG & EM \\
\hline  

150 & $\beta_0=-0.746$                                         & -0.045 & -0.063 & 0.222 & 0.434 \\
 &  $\beta_{1}=0.134$                                           & 0.007 & 0.012 & 0.030 & 0.058 \\
 &  $\gamma_1=0.316$                                          & -0.008 & 0.002 & 0.031 & 0.031 \\ 
 &  $\gamma_2=0.179$                                          & 0.001 & 0.003 & 0.013 & 0.013\\
 &  $\alpha=0.500$                                                & 0.024  & 0.039 & 0.155 & 0.365\\[0.5ex]
 
200  & $\beta_0=-0.746$                                         & -0.010 & -0.030 & 0.110 & 0.374 \\
  &  $\beta_{1}=0.134$                                           & 0.000 & 0.006 & 0.020 & 0.049 \\
  &  $\gamma_1=0.316$                                          & -0.005 & -0.002 & 0.029 & 0.027 \\ 
  &  $\gamma_2=0.179$                                          & 0.001 & 0.002 & 0.016 & 0.012\\
  &  $\alpha=0.500$                                                & 0.001  & 0.005 & 0.069 & 0.355\\[0.5ex]
 
 150   & $\beta_0=-0.692$                                       & -0.035 & -0.008 & 0.191 & 0.415 \\
  &  $\beta_{1}=0.156$                                           & 0.004 & 0.003 & 0.029 & 0.064 \\
  &  $\gamma_1=0.316$                                          & -0.010 & 0.001 & 0.031 & 0.032 \\ 
  &  $\gamma_2=0.179$                                          & 0.000 & 0.000 & 0.013 & 0.012\\
  &  $\alpha=0.750$                                                & 0.000  & -0.059 & 0.134 & 0.352\\[0.5ex]
 
 200  & $\beta_0=-0.692$                                        & -0.005 & -0.046 & 0.088 & 0.412 \\
  &  $\beta_{1}=0.156$                                           & 0.000 & -0.001 & 0.022 & 0.056 \\
  &  $\gamma_1=0.316$                                          & -0.010 & -0.005 & 0.031 & 0.028 \\ 
  &  $\gamma_2=0.179$                                          & 0.001 & -0.001 & 0.018 & 0.012\\
  &  $\alpha=0.750$                                                & -0.005  & -0.099 & 0.091 & 0.350\\
\hline
\end{tabular}
\label{table:NCG2}
\end{table}
\begin{table} [ht!]
\caption{Bias and RMSE of cured proportions for BCT model with binary covariate and comparison of NCG algorithm with EM algorithm} 
\centering
\begin{tabular}{ l l l l l l l }\\ 
\hline                                 
$\alpha$ & $n$ & $p_0$ &\multicolumn{2}{c}{Bias}  &\multicolumn{2}{c}{RMSE} \\  \cline{4-5}  \cline{6-7}
& & &NCG & EM & NCG & EM \\
\hline  

0.500 & 200  & $p_{01} = 0.400$                            & 0.006 & 0.001 & 0.048 & 0.062 \\
 & &  $p_{00} = 0.200$                                          & 0.004 & 0.000 & 0.033 & 0.062 \\[0.5ex]
 
0.500 & 300  & $p_{01} = 0.400$                          & -0.001 & 0.003 & 0.042 & 0.054 \\
 & & $p_{01} = 0.200$                                          & 0.000 & 0.002 & 0.026 & 0.050 \\[0.5ex]
 
0.750 & 200  & $p_{01} = 0.650$                            & 0.005 & -0.001 & 0.053 & 0.065 \\
 & &  $p_{00} = 0.350$                                           & 0.005 & -0.001 & 0.048 & 0.073 \\[0.5ex]
 
0.750 & 300  & $p_{01} = 0.650$                             & 0.006 & -0.003 & 0.036 & 0.052 \\
 & &  $p_{01} = 0.350$                                            & 0.006 & 0.001 & 0.027 & 0.059 \\[0.5ex]
\hline
\end{tabular}
\label{table:cure}
\end{table}
\begin{table} [ht!]
\caption{CPU times (in seconds) for NCG and EM algorithms}
\renewcommand{\arraystretch}{1.1}
\centering
\begin{tabular}{ l l l l}\\ 
\hline                                 
$n$ & $(\alpha,p_{01},p_{00})$ &\multicolumn{2}{c}{CPU Time (seconds)} \\  \cline{3-4} 
& &NCG & EM \\
\hline  
200 & $(0.500,0.400,0.200)$ & 119.62    &  83.97  \\[0.5ex]
300 & $(0.500,0.400,0.200)$ & 33.12    & 103.03   \\[0.5ex]
200 & $(0.750,0.400,0.200)$ & 96.12    &  68.89  \\[0.5ex]
300 & $(0.750,0.400,0.200)$ & 25.73    &  92.92  \\[0.5ex]
200 & $(0.750,0.650,0.350)$ & 301.19    &  84.03  \\[0.5ex]
300 & $(0.750,0.650,0.350)$ & 104.10    & 110.27   \\[0.5ex]
\hline
\end{tabular}
\label{table:CPU}
\end{table} 

\section{Real data analysis}

To illustrate the proposed NCG algorithm in the context of BCT cure rate model, we used the well-known cutaneous melanoma data (also used in Pal and Balakrishnan, 2018); see Ibrahim et al. (2001) for a description of the data. As done by Pal and Balakrishnan (2018), we considered the nodule category $(1: n_1=111; 2: n_2 = 137; 3: n_3 = 87;$ and $4: n_4 = 82)$ as the only covariate $(x)$ in our model. For computing the initial values of model parameters for the NCG algorithm, we first calculated the non-parametric estimates of the cured proportions for nodule categories 1 and 4 (obtained by taking the Kaplan-Meier estimate of the survival function at the largest observed lifetime). Then, we equated these to $p_0(x)$ with $x=1$ and 4, as in \eqref{p0}, for different values of $\alpha$ in the interval [0, 1]. By solving the two equations for different values of $\alpha$, we obtained a set of values of the regression parameters $\beta_0$ and $\beta_1$. For each $(\beta_0,\beta_1)$ and the corresponding value of $\alpha,$ we calculated the log-likelihood function value. For this purpose, we equated the mean and variance of the Weibull density function in \eqref{wei} to the mean and variance of the observed data to come up with a choice of the parameters $\gamma_1$ and $\gamma_2$. Finally, we selected that set of model parameters, as our initial guess, for which the log-likelihood value turned out to be the maximum. 

With the aforementioned technique of finding initial values, the NCG algorithm simultaneously maximized all model parameters. The estimates were obtained as $\hat{\beta_0} = -1.238,$ $\hat{\beta_1} = 0.358,$ $\hat{\gamma_1} = 0.585,$ $\hat{\gamma_2} = 0.393,$ and $\hat{\alpha} = 0.003$ with the maximized log-likelihood value as -513.125. As pointed out by Pal and Balakrishnan (2018), the EM algorithm did not allow simultaneous maximization of all model parameters and the authors proposed a profile likelihood technique within the EM algorithm. The corresponding EM estimates are $\hat{\beta_0} = -1.275,$ $\hat{\beta_1} = 0.374,$ $\hat{\gamma_1} = 0.578,$ $\hat{\gamma_2} = 0.393,$ and $\hat{\alpha} = 0.050$ (using profile likelihood) with the maximized log-likelihood value as -513.252. We note that the proposed NCG algorithm results in a better fit. To compute the standard errors by inverting the matrix of negative of the second-order derivatives of the log-likelihood function, we noted that the second order derivatives are highly unstable. As such, for the NCG algorithm, we recommend a bootstrap technique (for instance, a non-parametric bootstrap) to find the standard errors of the model parameters. We computed the non-parametric bootstrap estimates of the standard errors (s.e.), based on a bootstrap sample of size 500, as $\text{s.e}(\beta_0) = 0.114,$ $\text{s.e}(\beta_1) = 0.046,$ $\text{s.e}(\gamma_1) = 0.039,$ $\text{s.e}(\gamma_2) = 0.033,$ and $\text{s.e}(\alpha) = 0.034$.

\begin{figure}[ht!]
\centering
\includegraphics[width=1.0\linewidth]{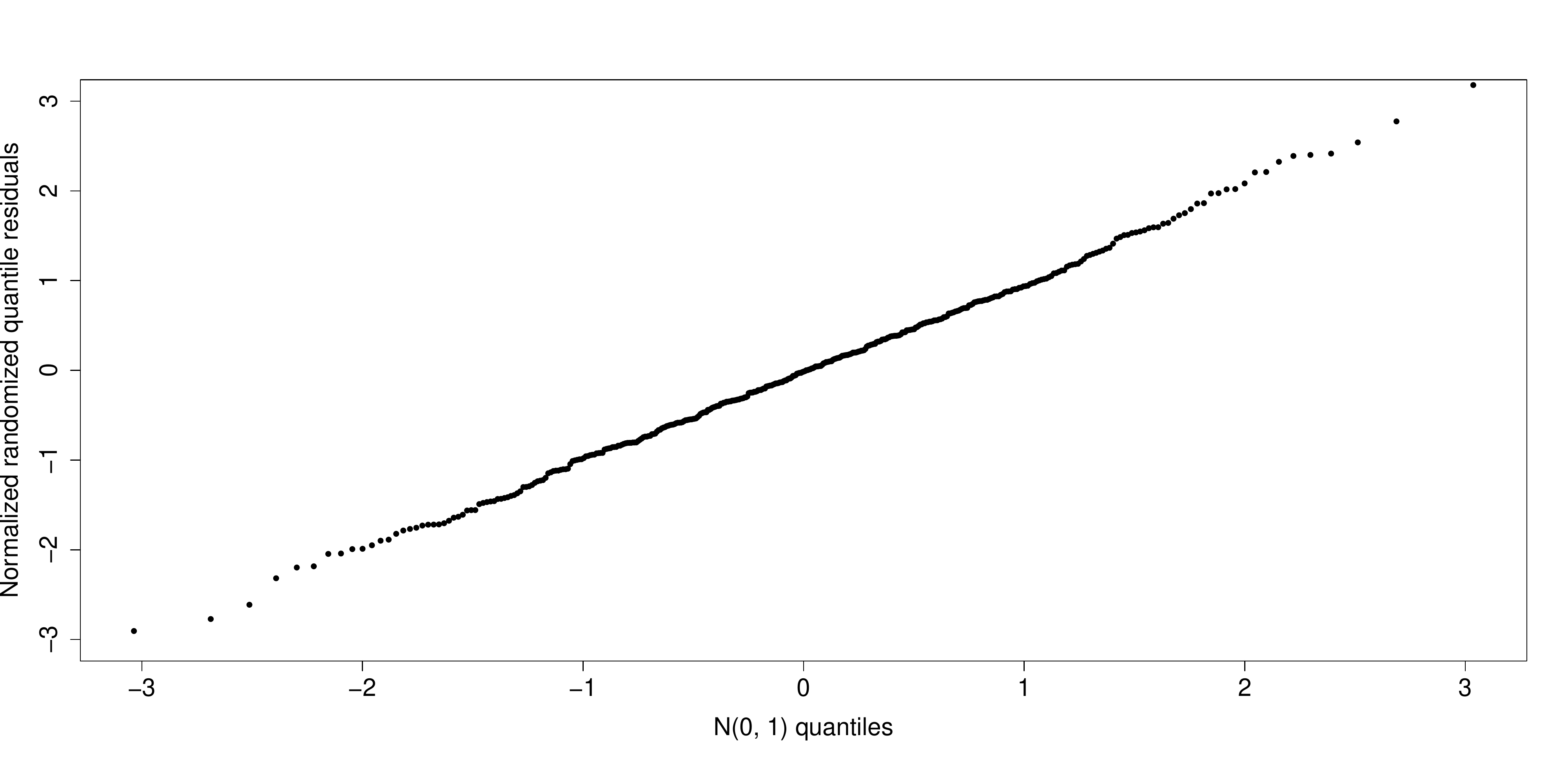}
\caption{QQ plot of normalized randomized quantile residuals}
\label{figure:QQ}
\end{figure}

To check for the adequacy of BCT model using NCG estimates, we computed the normalized randomized quantile residuals; see Dunn and Smyth (1996). In Figure \ref{figure:QQ}, we present the QQ plot where each point corresponds to the median of five sets of ordered residuals. The plot clearly suggests that the BCT cure rate model provides an adequate fit to the melanoma data. We also formally tested for the normality of residuals using Kolmogorov-Smirnov test and the p-value turned out to be 0.995, which provides strong evidence of normality of residuals.

\section{Concluding remarks}

In this paper, we have proposed a new algorithm for the maximum likelihood estimation of the BCT cure rate model parameters. We have compared the performance of our proposed NCG algorithm with the EM algorithm. Unlike the EM algorithm, the NCG algorithm allows simultaneous maximization of all model parameters. Furthermore, the NCG algorithm, when compared to the EM algorithm, results in a lower bias and a significantly lower RMSE of the parameters that are related to the cured proportion, i.e., $\beta_0, \beta_1,$ and $\alpha$. This ultimately results in a lower RMSE of the estimates of the cured proportions. We also compared the CPU times and have seen that for larger sample size, the NCG algorithm runs faster than the EM algorithm. The NCG algorithm may be useful for other complicated cure rate models, for instance, the Conway-Maxwell (COM) Poisson cure rate model (Balakrishnan and Pal, 2016) and the destructive COM-Poisson cure rate model (Pal et al., 2018), where a profile likelihood approach has been proposed for the estimation of the COM-Poisson shape parameter $\phi$ due to the flatness of the likelihood surface with respect to $\phi$. It will be interesting to see how the NCG algorithm performs in these cases and how it compares with the EM algorithm. We are currently working on these and hope to report the findings in a future paper. 

\section*{Conflict of Interest}
On behalf of all authors, the corresponding author states that there is no conflict of interest. 

\section*{Acknowledgment}

The first author express his thanks to University of Texas at Arlington, USA for providing the Research Enhancement Program (REP) fund to support this research.

\section*{References}

\begin{description}

\item  Adesokan, B., Knudsen, K., Krishnan, V. P. and Roy, S. (2018) A fully non-linear optimization approach to acousto-electric tomography. {\it Inverse Problems} {\bf 34}, 104004.

\item Annunziato, M. and Borz\`i, A. (2013) A Fokker–Planck control framework for multidimensional stochastic process. {\it Journal of Computational and Applied Mathematics} {\bf 237}, 487-507.

\item Balakrishnan, N., Barui, S., and Milienos, F. (2017). Proportional hazards under Conway–Maxwell-Poisson cure rate model and associated inference. {\it Statistical Methods in Medical Research} {\bf 26}, 2055$–$2077.

\item Balakrishnan, N., Koutras, M. V., Milienos, F. S., and Pal, S. (2016). Piecewise linear approximations for cure rate models and associated inferential issues. {\it Methodology and Computing in Applied Probability} {\bf 18}, 937$-$966.

\item Balakrishnan, N. and Pal, S. (2013). Lognormal lifetimes and likelihood-based inference for flexible cure rate models based on COM-Poisson family. {\it Computational Statistics \& Data Analysis} {\bf 67}, 41$-$67.

\item Balakrishnan, N. and Pal, S. (2016). Expectation maximization-based likelihood inference for flexible cure rate models with Weibull lifetimes. {\it Statistical Methods in Medical Research} {\bf 25}, 1535$-$1563.

\item Berkson, J. and Gage, R. P. (1952). Survival curve for cancer patients following treatment. {\it Journal of the American Statistical Association} $\boldsymbol{47}$, 501$-$515.

\item Boag, J. W. (1949). Maximum likelihood estimates of the proportion of patients cured by cancer therapy. {\it Journal of the Royal Statistical Society: Series B} $\boldsymbol{11},$ 15$-$53.

\item Box, G. E. P. and Cox, D. R. (1964). An analysis of transformations (with discussion). {\it Journal of the Royal Statistical Society: Series B} $\boldsymbol{26},$ 211$-$252.

\item Castro, M., Cancho, V. G. and Rodrigues, J. (2010). A note on a unified approach for cure rate models. {\it Brazilian Journal of Probability and Statistics} {\bf 24}, 100$-$103.

\item Chen, M. -H., Ibrahim, J. G. and Sinha, D. (1999). A new Bayesian model for survival data with a surviving fraction. {\it Journal of the American Statistical Association} $\boldsymbol{94}$, 909$-$919.

\item de Freitas, L. A. and Rodrigues, J. (2013). Standard exponential cure rate model with informative censoring. {\it Communications in Statistics - Simulation and Computation} {\bf 42}, 8$–$23.

\item Diao, G. and Yin, G. (2012). A general transformation class of semiparametric cure rate frailty models. {\it Annals of the Institute of Statistical Mathematics} $\boldsymbol{64},$ 959$-$989.

\item Dunn, P. K. and Smyth, G. K. (1996). Randomized quantile residuals. {\it Journal of Computational and Graphical Statistics} {\bf 5}, 236$–$244.

\item Farewell, V. T. (1986). Mixture models in survival analysis: are they worth the risk? {\it Canadian Journal of Statistics} {\bf 14}, 257$-$262.

\item Hager, W. W. and Zhang, H (2005) A new conjugate gradient method with guaranteed descent and an efficient line search. {\it SIAM Journal on Optimization} {\bf 16}, 170-192.

\item Ibrahim, J. G., Chen, M. -H. and Sinha, D. (2001). {\it Bayesian Survival Analysis}. New York: Springer-Verlag.

\item Kannan, N., Kundu, D., Nair, P. and Tripathi, R. C. (2010). The generalized exponential cure rate model with covariates. {\it Journal of Applied Statistics} {\bf 37}, 1625$-$1636.

\item Koutras, M. V. and Milienos, F. S. (2017). A flexible family of transformation cure rate models. {\it Statistics in Medicine} {\bf 36}, 2559$-$2575.

\item Kuk, A. Y. C. and Chen, C. H. (1992). A mixture model combining logistic regression with proportional hazards regression. {\it Biometrika} {\bf 79}, 531$-$541.

\item Li, C. S. and Taylor, J. M. G. (2002). A semi-parametric accelerated failure time cure model. {\it Statistics in Medicine} {\bf 21}, 3235$–$3247.

\item Maller, R. A. and Zhou, X. (1996). {\it Survival Analysis with Long-Term Survivors}. New York: John Wiley \& Sons.

\item Neittaanmaki, P. and Tiba, D. (1994). Optimal Control of Nonlinear Parabolic Systems: Theory, Algorithms and Applications. {\it Pure and Applied Mathematics. CRC Press, London}.

\item Pal, S. and Balakrishnan, S. (2018). Expectation maximization algorithm for Box–Cox transformation cure rate model and assessment of model misspecification under Weibull lifetimes. {\it IEEE Journal of Biomedical and Health Informatics} {\bf 22}, 926$-$934.

\item Pal, S., Majakwara, J. and Balakrishnan, N. (2018). An EM algorithm for the destructive COM-Poisson regression cure rate model. {\it Metrika} {\bf 81}, 143$-$171.

\item Peng, Y. and Xu, J. (2012). An extended cure model and model selection. {\it Lifetime Data Analysis} {\bf 18}, 215$-$233.

\item Rodrigues, J., de Castro, M., Cancho, V. G. and Balakrishnan, N. (2009). COM-Poisson cure rate survival models and an application to a cutaneous melanoma data. {\it Journal of Statistical Planning and Inference} $\boldsymbol{139},$ 3605$-$3611.

\item Roy, S., Annunziato, M. and Borz\`i, A. (2016). A Fokker–Planck feedback control-constrained approach for modelling crowd motion. {\it Journal of Computational and Theoretical Transport} {\bf 45}, 442-458.

\item Roy, S., Annunziato, M., Borz\`i, A. and Klingenberg, C. (2018). A Fokker–Planck approach to control collective motion. {\it Computational Optimization and Applications} {\bf 69}, 423-459.

\item Roy, S., Borz\`i, A. and Habbal, A. (2017). Pedestrian motion constrained by FP-constrained Nash games. {\it Royal Society Open Science}, {\bf 4}, 170648.

\item Sun, L., Tong, X. and Zhou, X. (2011). A class of Box-Cox transformation models for recurrent event data. {\it Lifetime Data Analysis} {\bf 17}, 280$-$301.

\item Yin, G. and Ibrahim, J. G. (2005). Cure rate models: a unified approach. {\it The Canadian Journal of Statistics} $\boldsymbol{33}$, 559$-$570.

\item Zeng, D., Yin, G. and Ibrahim, J. G. (2006). Semiparametric transformation models for survival data with a cure fraction. {\it Journal of the American Statistical Association} $\boldsymbol{101}$, 670$-$684.

\end{description}

\end{document}